\documentclass[a4paper,11pt]{article}
\usepackage{pos}
\usepackage{tikz}
\usetikzlibrary{external}
\newlength\figwidth
\newlength\figheight
\newcommand\mylabelsize\normalsize
\usepackage{pgfplots}
\usepackage{subcaption}
\usepackage{booktabs}
\usepackage{cleveref}
\usepackage{ifthen}
\usepackage{floatrow}
\newfloatcommand{capbtabbox}{table}[][\FBwidth]

\newcommand\unit[1]{\,\text{#1}}
\newcommand\XPT{$\chi$PT }
\newcommand\mPS{m_\text{PS}}
\newcommand\mPCAC{m_\text{PCAC}}
\newcommand\fPS{f_\text{PS}}

\newcommand\sym{\text{sym}}
\newcommand\cont{\text{cont}}
\newcommand\phys{\text{phys}}
\newcommand\sqt{\sqrt{t_0}}
\newcommand\fpik{f_{\pi K}}
\newcommand\tr[1]{\mathrm{tr}\left( #1 \right)}


\title{Scale Setting for CLS 2+1 Simulations}
\author*[a]{Ben Stra{\ss}berger}
\author[f] {Marco C\`e}
\author[c] {Sara Collins}
\author[g] {Antoine G\'erardin}
\author[b] {Georg von Hippel}
\author[e] {Piotr Korcyl}
\author[d] {Tomasz Korzec}
\author[h] {Daniel Mohler}
\author[a] {Andreas Risch}
\author[a] {Stefan Schaefer}
\author[c] {Wolfgang S\"oldner}
\author[a] {Rainer Sommer}

\affiliation[a]{John von Neumann Institute for Computing (NIC), Deutsches Elektronen-Synchrotron DESY,\\
Platanenallee 6, 15738 Zeuthen, Germany}

\affiliation[b]{Johannes Gutenberg-Universit\"at Mainz\\
Johann-Joachim-Becher Weg 45, 55099 Mainz, Germany}

\affiliation[c]{Institut f\"ur Theoretische Physik, Universit\"at Regensburg,\\
Universit\"atsstra{\ss}e 31, 93040 Regensburg, Germany}

\affiliation[d]{Bergische Universit\"at Wuppertal,\\
Gau{\ss}stra{\ss}e 20, 42119 Wuppertal, Germany}

\affiliation[e]{Institute of Theoretical Physics, Jagiellonian University,\\
ul. Gołębia 24, 31-007 Kraków, Poland}

\affiliation[f]{Department of Theoretical Physics, CERN,\\
1211 Geneva 23, Switzerland}

\affiliation[g]{Centre de Physique Théorique, Campus of Luminy,\\
Case 907, F-13288 Marseille cedex 9, France}
   
\affiliation[h]{Helmholtzzentrum für Schwerionenforschung (GSI),\\
Planckstrasse 1, 64291 Darmstadt, Germany}

\emailAdd{ben.strassberger@desy.de}
\emailAdd{stefan.schaefer@desy.de}

\abstract{We present an update of the scale setting for $N_f=2+1$ flavor QCD using gradient flow scales and pseudo-scalar decay constants. We analyze the latest ensembles with $2+1$ flavors of non-perturbatively improved Wilson fermions generated by CLS for improved precision. Special care is taken to correct for mistuning by measuring directly the mass derivatives of the various observables. We determine $t_0$ with input taken from a combination of leptonic decay rates of the Pion and the Kaon.
}

\renewcommand{\hookAfterAbstract}{%
\par\bigskip
\textsc{Preprint Number:}
DESY-21-175, WUB/21-05
}

\FullConference{%
 The 38th International Symposium on Lattice Field Theory, LATTICE2021
  26th-30th July, 2021
  Zoom/Gather@Massachusetts Institute of Technology
}

\begin{document}
\maketitle

\section{Introduction}
%\begin{itemize}
   %\item $N_f = 2+1$, non-perturbatively improved, mostly obc
   %\item lattice spacings from $0.037\,\mathrm{fm}$ to $0.085\,\mathrm{fm}$
   %\item pion masses from $430\,\mathrm{MeV}$ to $134\,\mathrm{MeV}$ (one ensemble at phys. pion mass)
   %\item
   %\item updated analysis from 2016~\cite{Bruno:2016plf}
   %\item combination of pion and kaon decay constants
   %\item flow scale $t_0$ as intermediate scale
   %\item average measurements from different groups (highly correlated)
   %\item fully correlated error analysis
   %\item include sign reweighting
%\end{itemize}

CLS is a consortium which has generated a set of gauge field configurations
with non-perturbatively improved Wilson fermions~\cite{Bruno:2014jqa,Mohler:2017wnb}. One of the basic tasks in
such an endeavor is the determination of the lattice spacing. Since the initial
analysis of the scale for the CLS $2+1$~flavor ensembles~\cite{Bruno:2016plf},
a much larger dataset has become available.
%\todo{CITE inital cls21 paper, improvement paper}

In this updated analysis, 20
ensembles with lattice spacings from $0.085\unit{fm}$ 
to $0.037\unit{fm}$ and Pion masses from $430\unit{MeV}$ to
$134\unit{MeV}$ are included. The scale is set using a combination of Pion and Kaon
decay constants and the flow scale $t_0$~\cite{Luscher:2010iy} as an intermediate scale. 
The value of this intermediate scale in physical units 
is relevant for the precision determination of
$\alpha_s$ \cite{Bruno:2017gxd},
the $\eta$ and $\eta^\prime$ masses and decay constants \cite{Bali:2021qem},
moments of distribution amplitudes \cite{RQCD:2019hps,RQCD:2019osh},
nucleon axial form factors \cite{RQCD:2019jai},
the proton radius \cite{Djukanovic:2021cgp},
the Muon magnetic moment \cite{Gerardin:2019rua},
and more.
%\todo{other CLS, if possible published }

As an improvement
with respect to the previous analysis, we also include the reweighting factors
originating from the negative sign of the strange quark determinant
\cite{Mohler:2020txx}, which occur on a small subset of the gauge field configurations.

Our ensembles have been generated along a line of constant sum  of the bare
quark masses $\tr{M} = m_u+m_d+m_s = \text{const}$. For each coupling, this sum has been
tuned such that this line approximately passes through the point of physical light and
strange mass. Of course, the precise value of this is only known after the analysis has
been completed. We therefore have to deal with a certain amount of mistuning, for which
we use the same method as in the previous analysis, i.e. by computing the derivatives of
our observables with respect to the quark masses.

We measure two-point correlators and extract the pseudo-scalar mass, $\mPS$, and
decay constant, $\fPS$, for the Pion and the Kaon as well as the PCAC mass,
$\mPCAC$, for the corresponding quark combinations. The extraction of these
quantities is done using plateau averages and fits as laid out in
\cite{Bruno:2016plf}. The gradient flow scale $t_0$ is defined by the clover definition of the action density and the Wilson flow~\cite{Luscher:2010iy}.
Its improved definition~\cite{Ramos:2015baa} was not yet available when
the simulations were planned. 
The measurements are then subjected to the next-to-leading order $\chi$PT finite volume correction according to
%\cite{Colangelo:2005gd,Colangelo:2003hf}.
%\cite{Gasser:1987zq}.
\cite{Gasser:1986vb}.
%\todo{add or replace by Gasser Leutwy}
%, which corrects our values by less than the statistical error.
The finite volume correction does not exceed the statistical error of the respective quantities.
We nevertheless include $50\%$ of the correction as an additional uncertainty.

%\todo{Define $t_0$; Clover action}
From these measurements we calculate the dimensionless quantities,
%\begin{align}
   %\phi_2 &= 8 t_0\,m_\pi^2  &
   %\phi_4 &= 8 t_0 \left( m_K^2 + \frac12 m_\pi^2 \right) \label{eq:phi4}\\
   %\sqt \fpik &= \sqt \frac23 \left( f_K + \frac12 f_\pi \right). \label{eq:sqrtt0_fpik}
%\end{align}
\begin{equation}
   \phi_2 = 8 t_0\,m_\pi^2, \quad
   \phi_4 = 8 t_0 \left( m_K^2 + \frac12 m_\pi^2 \right), \label{eq:phi2_phi4}
\end{equation}
\begin{equation}
   \sqt \fpik = \sqt \frac23 \left( f_K + \frac12 f_\pi \right). \label{eq:sqrtt0_fpik}
\end{equation}

The combination of Pion and Kaon decay constants $\fpik$ will be used to set the scale. 
$\phi_2$ and $\phi_4$ will form the basis of the analysis since in lowest order $\chi$PT $\phi_2 \propto m_u+m_d$ and $\phi_4 \propto m_u+m_d+m_s$. 

In our analysis, we define lines of constant physics by setting $\phi_4$ to 
a certain value.
We use $\phi_4$, because it is non perturbatively improved up to order $a^2$ along the CLS quark-mass trajectory.

We therefore proceed with the following steps: by  measuring
$\sqt \fpik$ and its mass derivatives on each ensemble, we can 
predict $\sqt \fpik$ at  given values of $\phi_4,\phi_2$. The chiral behavior of this 
data can now be fitted and taken to the continuum. By tuning $\phi_4$ such
that the continuum curve passes through the physical point, we can then 
determine the physical value of $t_0$ and the lattice spacing at which we had done 
the simulations.

%\section{Lattice Measurements}
%\begin{itemize}
   %\item list ensembles? show plot of ensemble landscape?
   %\item explain ensembles and trajectories
   %\item introduce correlators
   %%\item explain plateaus from excited state fit
   %%\item explain primary measurements $t_0, \mPS, \mPCAC, \fPS$
   %\item define secondary measurements $\phi_2, \phi_4, \sqt \fpik$ (dimensionless)
%\end{itemize}

%\setlength\figwidth{0.6\textwidth}
%\setlength\figheight{0.66\figwidth}
%\renewcommand\mylabelsize\normalsize
%\begin{figure}[pbt]
   %\centering
   %\input{../../ana_bstrass/figures/tikz/phi2_t0inv.tikz}%
   %\caption{Ensemble landscape}
   %\label{fig:ensembles}
%\end{figure}

\section{Observables at physical $\phi_4$}
\begin{figure}[pbt]
   \centering
   \includegraphics{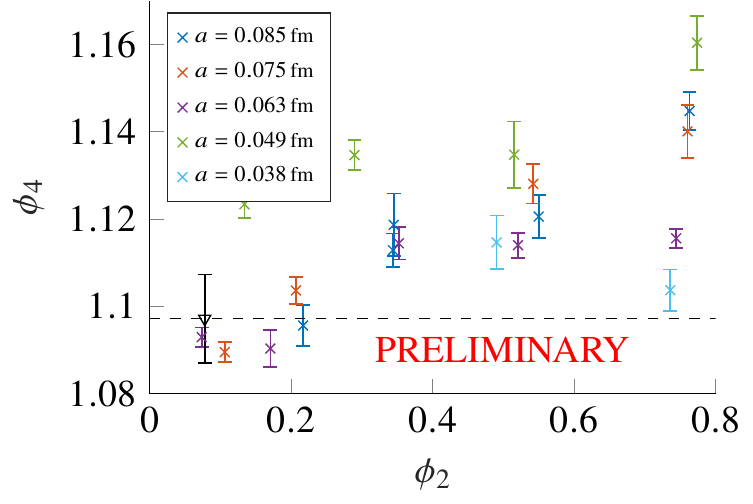}%
   \caption{Measurements for $\phi_4$ show the mistuning in $\phi_4=\text{const}$. The physical point, as calculated by this analysis, is shown as a triangle.}
   \label{fig:phi2_phi4}
\end{figure}

Since the ensembles lie on a line of constant sum of the bare quark masses, more precisely
$ 2/\kappa_l + 1/\kappa_s = \mathrm{const}$,
the  condition $\phi_4=\mathrm{const}$ is certainly not fulfilled for all ensembles due to 
discretization effects and higher order effects in \XPT. On top of that
comes the fact that the physical  $\phi_4$ has not been known during
the planning of the simulations --- and also depends on the particular discretization chosen for $t_0$.
In \cref{fig:phi2_phi4}, we present our ensembles in the $\phi_2$--$\phi_4$ plane
and observe that the sum of the three quark masses is within $8\%$ of the physical value.
\setlength\figwidth{0.4\textwidth}%
\setlength\figheight{0.66\figwidth}% 
\renewcommand\mylabelsize\tiny%

To get the observables at a given value of $\phi_4$, we measured the
derivatives $\frac{dX}{dm_i}$ of the observables~$X$ with respect to the quark masses of all the involved measurements.
 With these we can construct the derivatives with respect to
$\phi_4$,
\begin{equation}
   \frac{dX}{d\phi_4} = \sum_{i=1}^3 n_i \frac{dX}{dm_i}/\left( \sum_{i=1}^3 n_i \frac{d\phi_4}{dm_i} \right).
   \label{eq:dXdphi4}
\end{equation}
Here $\vec{n}$ is the direction of the shift in the space of quark masses.
%The direction $\vec{n} = \left\{ n_i \right\}$ in the space of the quark masses
It has an effect on distance of the shift needed to reach the given value of
$\phi_4$ and thus on the resulting uncertainty.
%\todo{grammar?}
For the symmetric ensembles we use $\vec{n} =
(1,1,1)/\sqrt{3}$ to preserve the symmetry and which was also the choice in
\cite{Bruno:2016plf}. For other ensembles, however, we use the direction $\vec n = (0,0,1)$,
which we found to be close to optimal, in the sense that it minimizes the errors of the shifted values of $f_{\pi K}$.
Since the shifts are typically less than $5\%$ in the sum
of the quark masses, we do expect the leading order of the Taylor expansion to
give results with a systematic error  below our statistical uncertainty.
This  also has been verified with a few ensembles at the symmetric line at a
different sum of quark masses.

In the 2016 analysis \cite{Bruno:2016plf}, we shifted the results on each ensemble individually. In our update, we now model
the mass derivatives of the observables as a function of quark mass and lattice spacing. This makes the
predictions more stable and also allows the  use of  ensembles, where the derivatives 
have not been measured.
With these derivatives the measurements can now be shifted to the desired~$\phi_4$ by
\begin{equation}
   X(\phi_4^\phys) = X(\phi_4) + \frac{dX}{d\phi_4} \left( \phi_4^\phys - \phi_4 \right).
   \label{eq:Xshift}
\end{equation}

\section{Chiral and Continuum Extrapolation}

In the description of the analysis, we now have data at any given value of $\phi_4$, 
which is a proxy for the sum of the two degenerate light and strange quark masses. To get to the physical point, 
we need to describe its chiral behavior and extrapolate to physical  light quark masses given by $\phi_2^\phys$. 

Chiral perturbation theory \cite{Gasser:1984gg,Bar:2013ora} predicts  
%\todo{Richtiges Paper von Baer zitiert?\cite{Bar:2013ora}}
\newcommand{\fbar}{\bar F}
\begin{equation}
   \begin{aligned}
      \fpik &= f \left[ 1 - \frac76 L\left(\frac{\phi_2}{\fbar^2}\right) - \frac43 L\left(\frac{\phi_4-\frac12 \phi_2}{\fbar^2}\right) - \frac12 L\left(\frac{\frac43 \phi_4-\phi_2}{\fbar^2}\right) + k \phi_4 +\mathcal{O}\left(M^2\right) \right]
   %&=: F_{\chi\text{PT}}(\phi_2,\phi_4) 
   \end{aligned}
   \label{eq:fpik_XPT}
\end{equation}
for the quark mass dependence of $\fpik$ in terms of a single SU(3) $\chi$PT NLO low energy constant, $k\propto (L_5+3L_4)$ as well as
\begin{equation}
   L(x) = x \log(x) \,,\quad \fbar = 4\pi \sqrt{8t_0}\, f.
   %= 4\pi \sqrt{8t_0}\, \fpik + \mathrm{O}(M)\,. 
\end{equation}
%\todo{remove $fpik$ term}
We defined $\fbar$ in terms of $f$, the decay constant in the chiral limit.
%, but it can be expressed in terms
%of $\fpik$ at the considered order of $\chi$PT as indicated. 
We then use a fit function for the chiral and continuum\footnote{We note that logarithmic corrections 
of the $a^2$ terms in the form $a^2 \alpha_s(1/a)^{\hat \Gamma}$ are present~\cite{Husung:2021mfl}. The known leading exponent $\hat \Gamma$ is reasonably small and in particular 
we here use the Gradient Flow observable $t_0$. In this case
the leading $\hat \Gamma$ vanishes in the pure gauge theory~\cite{H:inprep2} and is not yet known in full QCD. We therefore ignore the presence of the logarithmic corrections at present.}
behavior 
\begin{align}
   %\begin{split}
   %F_\chi^\cont(\phi_2,\phi_4) = A &\left[ 1 - \frac76 L\left(\phi_2/\left( 8 (4\pi)^2 A^2\right)\right) - \frac43 L\left(\left(\phi_4-\frac12 \phi_2\right)/\left(8(4\pi)^2 A^2\right)\right) \right.\\
   %&\hphantom{\left[ 1 \right.}\left.- \frac12 L\left(\left(\frac43 \phi_4-\phi_2\right)/\left( 8(4\pi)^2 A^2\right)\right)  +B \phi_4 \right]    \label{eq:FTXPT}
   %\end{split}
   F_\chi(\phi_2,\phi_4) &= F_\chi^\cont(\phi_2,\phi_4) \cdot \left( 1 + C \cdot \frac{a^2}{t_0} \right) \label{eq:FTXPT_cont}\\
   F_\chi^\cont(\phi_2,\phi_4) &= \frac{A}{8\pi\sqrt{2}} \left[ 1 - \frac76 L\left(\frac{\phi_2}{A^2}\right) - \frac43 L\left(\frac{\phi_4-\frac12 \phi_2}{A^2}\right) - \frac12 L\left(\frac{\frac43 \phi_4-\phi_2}{A^2}\right)  +B \phi_4 \right]    \label{eq:FTXPT}
\end{align}
%\todo{$F_1 \rightarrow F_\chi$}
with the parameters $A=A(\phi_4)$ and $B=B(\phi_4)$ for each fixed value of $\phi_4$. 

\begin{figure}[pbt]
   \centering
   \begin{subfigure}[b]{0.49\textwidth}
      \centering
      \includegraphics{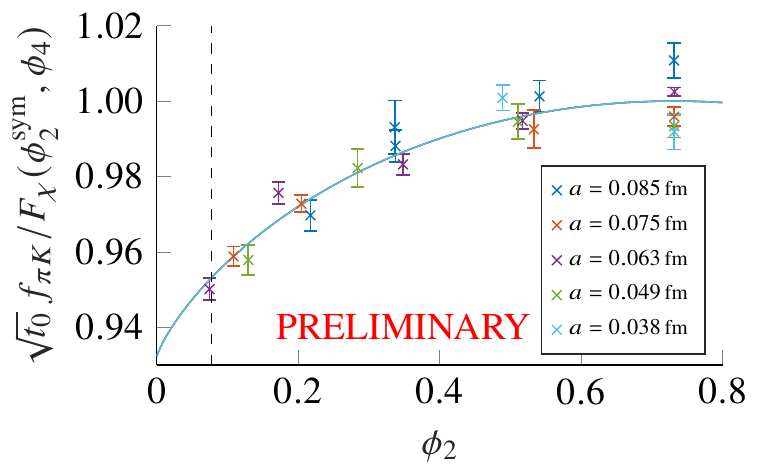}%
   \end{subfigure}
   \begin{subfigure}[b]{0.49\textwidth}
      \centering
      \includegraphics{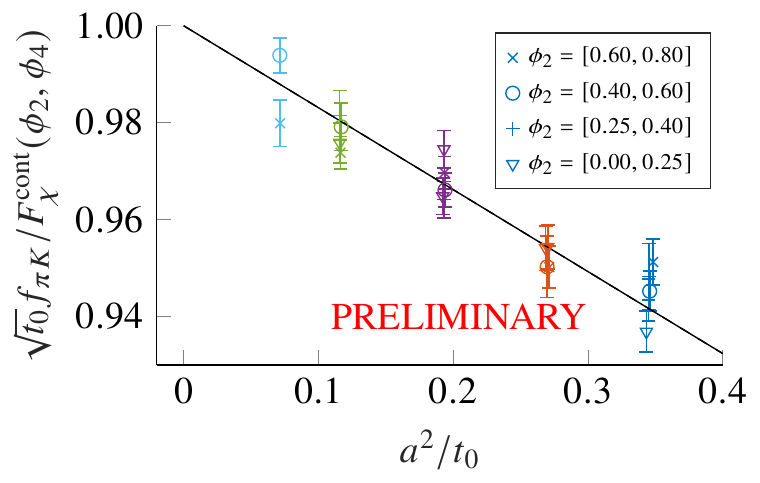}%
   \end{subfigure}
   \caption{The chiral extrapolation on the left shows the measured data of $\sqrt{t_0} \fpik$ normalized by the fit function at the symmetric point.
      The solid line is the NLO \XPT prediction, which does not depend on any NLO parameters and only logarithmically on the LO parameter $A$.
      No systematic deviation from this continuum formula can be detected.
   On the right, the continuum extrapolation of the same data, normalized by the fit-function evaluated for $a=0$, is shown.}
   \label{fig:extrap}
\end{figure}

It is worthwhile to consider the ratio of the chiral function
\begin{equation}
   \frac{F_\chi^\cont(\phi_2,\phi_4)}{F_\chi^\cont(\phi_2^\sym,\phi_4)}    
   = R_\chi(\phi_2,\phi_4) + \mathrm{O}(M^2)
   \label{eq:XPTratio} 
\end{equation}
since
\begin{equation}
 %R_\chi(\phi_2,\phi_4) =  1 - \frac76 \left( L_\pi-L_\pi^\sym \right) - \frac43 \left( L_K - L_K^\sym \right) - \frac12 \left( L_\eta - L_\eta^\sym \right)
   \begin{split}
   R_\chi(\phi_2,\phi_4) =  1 
   &- \frac76 L\left( \frac{\phi_2}{A^2} \right) 
   - \frac43 L\left( \frac{\phi_4-\frac12\phi_2}{A^2} \right) 
   - \frac12 L\left( \frac{\frac43\phi_4-\phi_2}{A^2} \right) \\
   &+ \frac76 L\left( \frac{\phi_2^\sym}{A^2} \right)
   + \frac43 L\left( \frac{\phi_4-\frac12\phi_2^\sym}{A^2} \right) 
   + \frac12 L\left( \frac{\frac43\phi_4-\phi_2^\sym}{A^2} \right)
   \end{split}
\end{equation}
%\todo{replace by L(...)}
 %is  determined by chiral perturbation theory without any free parameter.
is free of parameters in NLO $\chi$PT, except for a weak dependence on $A$ in the logarithms.
We observe no systematic deviations from NLO \XPT as shown in  \cref{fig:extrap}.
\setlength\figwidth{0.4\textwidth}%
\setlength\figheight{0.57\figwidth}%
\renewcommand\mylabelsize\tiny%

In the same spirit the discretization effects are illustrated in the right hand plot of the same figure.  Dividing each data point
by the continuum \XPT formula \cref{eq:FTXPT} we expect the ratio
\begin{equation}
   R_\cont = \frac{F_\chi(\phi_2,\phi_4)}{F^\cont_\chi(\phi_2,\phi_4)} = 1 + C\cdot \frac{a^2}{t_0}
   \label{eq:cont_ratio}
\end{equation}
to be a linear function in $a^2$ to leading order of the Symanzik expansion. Again, no deviation due to
higher order terms is detected.
%These considerations lead us to the  parametrization 
%\begin{equation}
   %F_\chi(\phi_2,\phi_4) = F_\chi^\cont(\phi_2,\phi_4) \cdot \left( 1 + C \cdot \frac{a^2}{t_0} \right)
   %\label{eq:continuum_extrap}
%\end{equation}
%of the data, from which we 
%extract the central value and statistical uncertainty
%of $\fpik\sqt$.

\begin{figure}[pbt]
   \centering
   \includegraphics{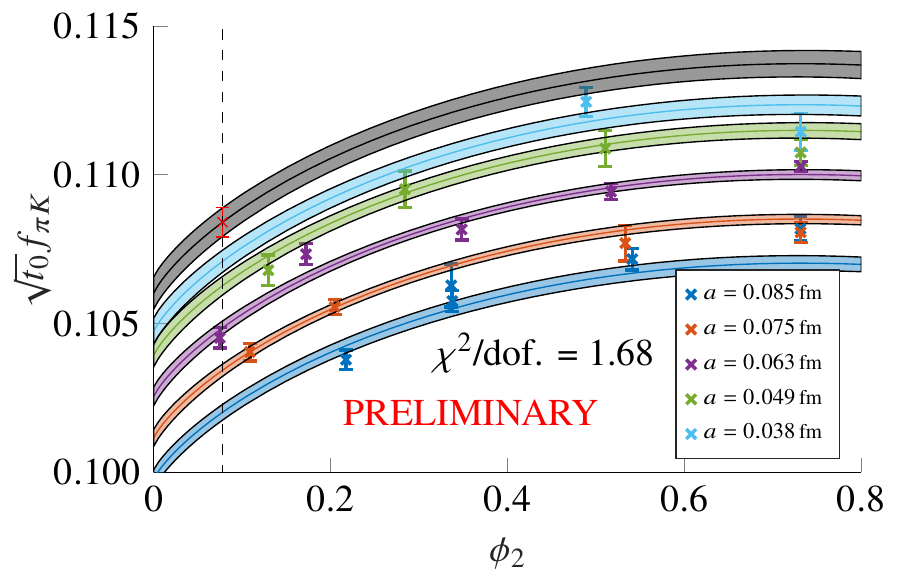}%
   \caption{Chiral and Continuum Extrapolation of $\sqrt{t_0} f_{\pi K}$.}
   %\todo{add line for $\beta=3.4$}
   \label{fig:phi2_sqrtt0_fpik}
\end{figure}

These considerations confirm the fit function given in \cref{eq:FTXPT_cont}, from which we extract the central value and statistical uncertainty of $\sqt\fpik$.
This central fit is displayed in \cref{fig:phi2_sqrtt0_fpik}.
It takes into account the data of the ensembles with $a<0.085\unit{fm}$. The stability when applying other cuts to the data is considered  later.
\setlength\figwidth{0.5\textwidth}%
\setlength\figheight{0.6\figwidth}%
\renewcommand\mylabelsize\tiny%

The combined chiral and continuum extrapolation is now evaluated at $\phi_2^\phys$
to determine $\sqt\fpik$ at the physical point.
Together with the values~\cite{Aoki:2021kgd,Zyla:2020zbs}
\begin{equation}
   f_\pi^\text{isoQCD} = 130.56(02)(13)(02)\unit{MeV}, \quad f_K^\text{isoQCD} = 157.2(2)(2)(4)\unit{MeV}
   \label{eq:f_exp}
\end{equation}
in isospin symmetric pure QCD
%\todo{explain errors (exp, QED, Vus)?? $\rightarrow$ Rainer?}
%\todo{experimental values $\rightarrow$ pure QCD\ldots}
we are able to extract
\begin{equation}
   \sqrt{t_0^\phys} = \frac{\sqt \fpik}{\fpik^\text{isoQCD}} = 
   0.1443(7)
\unit{fm} \quad \left(\chi\mathrm{PT\ fit}, a<0.08\unit{fm}\right)
   \label{eq:sqrtt0_phys_FTXPT}
\end{equation}
from the $\chi$PT fit. 
The isoQCD values in \cref{eq:f_exp} are obtained from the experimental decay rates of $\pi \to \ell \nu +\gamma$ and
$K \to \ell \nu +\gamma$ by taking out the QED and isospin effects with the help of \XPT. This introduces the 
largest (second) error, while the first error is due to the experimental decay rate, and the third is due to the uncertainty of the CKM matrix elements $V_{ud}$ and $V_{us}$.

%\todo{fixpoint iteration}
The physical scale $t_0^\phys$ enters in the beginning of the analysis to define the physical point $\left( \phi_2^\phys, \phi_4^\phys \right)$. We therefore find the fixpoint such that the resulting $t_0^\phys$ is the same one that is used in the definition of the physical $\phi_2$ and $\phi_4$.
For the statistical error of this final result the full correlation of the errors between the various
observables is taken into account. We find the physical point at
\begin{equation}
   \phi_2^\phys = 0.0779(7), \quad \phi_4^\phys = 1.098(10).
   \label{eq:phi2_phi4_phys}
\end{equation}

The fact that we do not observe significant deviations from the fit formula does, 
of course, not mean that they are not present.
To estimate this source of systematic error, we use a range of different extrapolations: 
the  \XPT formula has been substituted by a Taylor expansion in the quark masses around the symmetric point and
we also augmented the continuum extrapolation by an $a^2 m_\pi^2$ term. Applying a series of
cuts to the data by removing the coarsest lattices or the ones with larger pion masses, leads
also to valid description of the data.
All fits which we consider render a  $\chi^2/\mathrm{dof}$  between $1$ and $2$.
Fits where the probability to find a $\chi^2$ greater than the measured one is less than $5\%$ are discarded.
%\todo{use gof as criterium; Cut values with $\mathrm{gof}<0.1$}
We then take the minimum and the maximum central values and
use half their difference as our preliminary systematic error and arrive at
\begin{equation}
   \sqrt{t_0^\phys} =
   0.1443(7)(13)
   \unit{fm}.
   \label{eq:sqrtt0_phys_mean}
\end{equation}
At present, the systematic error dominates.

\section{Lattice Spacing}
\begin{figure}
   \begin{floatrow}
      \ffigbox{%
         \centering
         \hspace*{-0.3cm}
         \includegraphics{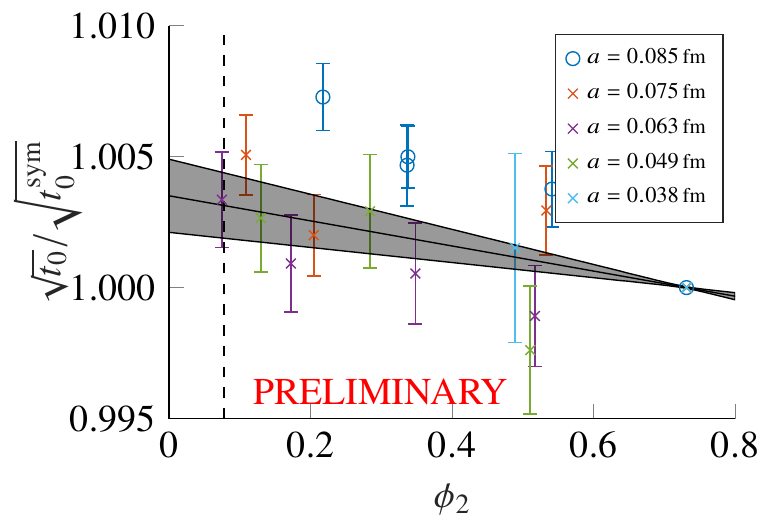}%
      }{%
         \caption{Scale $\sqt$ as a function of $\phi_2$ normalized by the scale at the symmetric point for each lattice spacing. The fit includes all but the coarsest ensemble with $a = 0.085\unit{fm}$.}
         \label{fig:phi2_sqrtt0}
      }
      \hspace*{2cm}
      \capbtabbox{
         \centering
         \begin{tabular}{cc}
            \toprule
            $\beta$ & $a\,\mathrm{[fm]}$\\
            \midrule
            3.40 &  0.0849(5)(8)\\
            3.46 &  0.0749(4)(7)\\
            3.55 &  0.0633(4)(6)\\
            3.70 &  0.0491(3)(4)\\
            3.85 &  0.0385(2)(3)\\
            \bottomrule
         \end{tabular}
      }{%
         \caption{Lattice spacings $a$ for each inverse coupling $\beta$ with statistic and systematic errors.}
         \label{tab:alat}
      }
   \end{floatrow}
\end{figure}

Having determined the intermediate scale $t_0^\phys$ at the physical point, we use it together with measurements for $t_0/a^2$ to calculate the lattice spacing $a$ in physical units.
Since measurements at the physical point are not available for all lattice spacings, we need to model the behavior of $t_0$ as a function of $\phi_2$.
Using next-to-leading order $\chi$PT~\cite{Bar:2013ora} we arrive at the fit formula
\begin{equation}
   R_{t_0}(\phi_2) = \frac{\sqrt{t_0}}{\sqrt{t_0^\sym}} = \sqrt{1 + G \left( \phi_2 - \phi_2^\sym \right)}.
   \label{eq:t0ratio_fit}
\end{equation}
The data along with the fit are shown in \cref{fig:phi2_sqrtt0}.
We see quite clearly that at $a=0.085\unit{fm}$ there are lattice artifacts which then disappear very quickly (more quickly than $a^2$).
This phenomenon was also observed in fit 4. of Ref.~\cite{Bruno:2016plf}.
We therefore perform a fit to the normalized scale $\sqrt{t_0/t_0^\sym}$ leaving out the coarsest ensembles with $a=0.085\unit{fm}$, as we have already done above. 
The figure is also good evidence for the smallness of 
mass-dependent $a^2$ effects. Further evidence is that the 
coefficients of mass-dependent cutoff effects in the Symanzik effective theory are very small for our discretization~\cite{Husung:2021mfl}.
\renewcommand\mylabelsize\tiny%
\setlength\figwidth{0.4\textwidth}%
\setlength\figheight{0.66\figwidth}%

The ratio $R_{t_0}(\phi_2^\phys)$ allows us to determine the scale $t_0$ in physical units at the symmetric point, $\sqrt{t_0^\phys}$ where we can use the measured values for $t_0/a^2$ to extract the lattice spacing, 
\begin{equation}
   a = \frac{\sqrt{t_0^\phys}}{R_{t_0}(\phi_2^\phys) } \cdot \frac{1}{\sqrt{t_0^\sym/a^2}}.
   \label{eq:alat}
\end{equation}
Using \cref{eq:sqrtt0_phys_mean} we arrive at the lattice spacings listed in \cref{tab:alat}.
%\begin{table}
   %\centering
   %\begin{tabular}{rccccc}
      %\toprule
      %\input{\filesdir/alat_table_flat.txt}%
      %\bottomrule
   %\end{tabular}
   %\caption{Lattice spacings $a$ for each inverse coupling $\beta$ with statistic and systematic errors.}
   %\label{tab:alat}
%\end{table}

\section{Conclusion and Outlook}
%\begin{itemize}
   %\item one method of many
   %\item advantages: raw data easy to determinae on the lattice
   %\item disadvantages: exp. uncert. of fK low; $V_{us}$ needed
   %\item compare to 2016: difference; less ``responsibility'' of the D200 lattice (best point in 2016); large fluctuation up
   %\item correction for isospin and QED; but not full theory; dependency on input ($\Omega$ mass,\ldots)
%\end{itemize}
The scale setting method presented here is one of many choices. Using the pseudo-scalar decay constants has the advantage that they can be easily and precisely calculated on the lattice. Contaminations by excited state contributions can be thoroughly controlled. On the other hand our method is limited by the necessity to relate the experimental decay rates which include photons in the final state to the pure QCD decay constants as well as the dependency on the CKM matrix element $V_{us}$. The latter means in particular that the validity of the Standard model at low energies is assumed. However, the estimated uncertainties due to QED and the CKM matrix elements are still significantly below our overall precision and  we are able to improve the result from the 2016 analysis~\cite{Bruno:2016plf}.
\setlength\figwidth{0.4\textwidth}%
\setlength\figheight{0.6\figwidth}%
\renewcommand\mylabelsize\normalsize%
\nocite{Bornyakov:2015eaa}%
\nocite{RBC:2014ntl}%
\nocite{Borsanyi:2012zs}%
\nocite{ExtendedTwistedMass:2021qui}%
\nocite{Miller:2020evg}%
\nocite{MILC:2015tqx}%
\nocite{Dowdall:2013rya}%
\begin{figure}[pbt]
   \centering
   \includegraphics{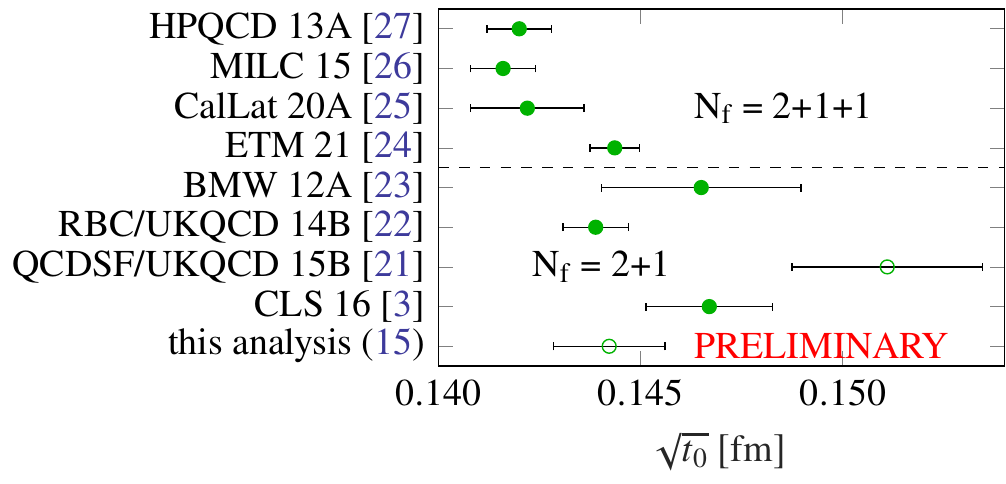}%
   \caption{Comparison of $\sqrt{t_0}$ for different groups. Open symbols data points are not published.}
   \label{fig:t0_ref_cmp}
\end{figure}
%\phantom{\cite{deDivitiis:2019xla}}
%\todo{Add BMW 20 value; only $w_0$ in~\cite{Aoki:2021kgd}}
\Cref{fig:t0_ref_cmp} compares the results from this analysis to previous determinations of $\sqrt{t_0}$ for $N_f = 2+1$ flavor and $N_f=2+1+1$ flavor ensembles. 
It is worth noting that the central value of the previous CLS determination (labeled CLS 16) is more
than $1\sigma$ above the current result. With the addition of several ensembles
close to the physical point, it can now be seen that the point closest to the
physical line in 2016 (purple point for $a = 0.063\unit{fm}$ at
$\phi_2\approx 0.17$ in \cref{fig:phi2_sqrtt0_fpik}) has a high statistical
fluctuation upwards. This resulted in the previous analysis being skewed.
It also highlights that precision scale setting, which is essential to precision results
from lattice QCD, is a challenging endeavor. We need large statistics such that autocorrelations
are under control as well as  data at a large range of lattice spacings
close to the continuum and quark masses sufficiently close to their physical values.
Going significantly beyond the present accuracy will also require an improved control of isospin breaking and QED effects.
%\\[1ex]
%{\bf Acknowledgments.}
\paragraph{Acknowledgments.}
S. Collins and R. Sommer were supported by the European Union’s Horizon 2020 research and innovation programme under the Marie Sk\l{}odowska-Curie grant agreement nos. 813942 (ITN EuroPLEx) and 824093 (STRONG- 2020).
We thank our colleagues in the Coordinated Lattice Simulations (CLS) effort 
[\url{http://wiki-zeuthen.desy.de/CLS/CLS}] for the joint generation of the gauge field ensembles on which the computation described here is based.

 We acknowledge PRACE for awarding us access to resource FERMI based in Italy at
CINECA, Bologna and to resource SuperMUC based in Germany at LRZ, Munich.
%We  are grateful for the support received by the computer centers.
We acknowledge the Gauss Centre for Supercomputing e.V. (\url{www.gauss-centre.eu})
for providing computing time through the John von Neumann Institute for
Computing (NIC) on JUQUEEN at J\"ulich
Supercomputing Centre and on  SuperMUC-NG at Leibniz Supercomputing Centre (\url{www.lrz.de}). We thank DESY for computing resources on the PAX cluster in Zeuthen.

%In the near future the systematic uncertainties will be studied further and we will examine the possibility to extract the scale separately from $f_\pi$ and $f_K$ as well as use the improved Zeuthen Flow~\cite{Ramos:2015baa}.

%\section*{Acknowledgments}
%The authors were supported by the European Union’s Horizon 2020 research and innovation programme under the Marie Skłodowska-Curie grant agreement nos. 813942 (ITN EuroPLEx) and 824093 (STRONG- 2020).
%S. Collins, W. S\"oldner and R. Sommer were supported by the European Union’s Horizon 2020 research and innovation programme under the Marie Sk\l{}odowska-Curie grant agreement nos. 813942 (ITN EuroPLEx) and 824093 (STRONG- 2020).

\bibliographystyle{JHEP}
%\bibliographystyle{plainnat}
%\bibliography{../../../literature/bibliography.bib}
\bibliography{./bibliography.bib}

\end{document}